\documentclass[a4paper,11pt]{article}

\usepackage{letter}
\usepackage{colortbl}
\usepackage{soul}
\usepackage{xltxtra} 
\usepackage{xgreek} 
\usepackage{fontspec}
\usepackage{listings}
\usepackage{graphicx}
\usepackage{fancyhdr}

\usepackage[english]{babel}
\usepackage{float}

\usepackage{colortbl}
\usepackage[table]{xcolor}
\definecolor{lightgray}{gray}{0.9}

\usepackage[margin=2.5cm]{geometry}

\usepackage[vlined,linesnumbered]{algorithm2e}

\SetKwInOut{Input}{Input}
\SetKwInOut{Output}{Output}
\SetKwInOut{Parameter}{Parameter}
\SetKwInput{Algorithm}{Algorithm}
\SetKwInput{Remark}{Remark}
\SetKw{Let}{Let}
\SetKw{Add}{Add}
\SetKwComment{Comment}{{// }}{}

\usepackage[backend=biber, style=numeric, sorting=none]{biblatex}
\definecolor{customgreen}{rgb}{0,0.6,0}
\definecolor{customgray}{rgb}{0.5,0.5,0.5}
\definecolor{custommauve}{rgb}{0.6,0,0.8}
\lstnewenvironment{SQLcode}{%
\lstset{language=SQL,%
showspaces=false,%
showstringspaces=false,%
tabsize=4,%
lineskip=2pt%
}}
{}

\lstnewenvironment{latex}{%
\lstset{language=TeX,%
keywordstyle=\bf\sffamily\color{blue},%
commentstyle=\color{ForestGreen},%
stringstyle=\color{Maroon},%
basicstyle=\ttfamily\normalsize\color{black},%
showspaces=false,%
showstringspaces=false,%
tabsize=4,%
aboveskip=10pt,%
belowskip=10pt,%
lineskip=2pt%
}}
{}

\usepackage{float}
\usepackage{amsmath}
\usepackage{fontspec}
\title{
\huge Simple Power Analysis on Post-Quantum Code Based Cryptosystems
}

\author{Konstantinos Spalas, MSc in Computer Science, Tripoli, dit2318cst@go.uop.gr \\
}

\date{3 June 2026}

\begin{document}

\maketitle


%
%
\begin{center}
  \section*{Abstract} 
\end{center}
Post-Quantum cryptography is about to substitute current cryptographic schemes as being resilient in attacks from quantum computers. McEliece and Bit Flip Key Encapsulation (BIKE) are two notable representatives based on coding theory, where classical structural attacks against these algorithms can be successfully phased out by selecting the appropriate key size. Using low cost equipment, the method of Simple Power Analysis (SPA) is used in this paper to evaluate whether there is information leakage during the decapsulation phase, where the shared secret key is generated. Executing a related experiment, it is shown that there is a significant correlation between electromagnetic emissions and secret values. In the aftermath, with only 250 power traces collected, machine learning models can predict a low sequence of secret bits of the shared session key, produced during the decapsulation.

\begin{center}
\[\]
\textbf{Key words} : Post-Quantum, Side-Channel Attacks, Simple Power Analysis 
\end{center}
%
%
\section{Introduction}\label{sec:intro}
Simple Power analysis (SPA) \cite{myors2010statistical} refers to a class of Side Channel Attacks (SCA) that exploit unintended information leakage produced during the execution of cryptographic algorithms. Rather than targeting only mathematical weaknesses in the cryptographic algorithm itself (structural attacks), SCA exploit observable characteristics such as power consumption, electromagnetic emissions, or execution timing to infer sensitive internal states. These leakages arise from the physical implementation of a system and are often influenced by hardware architecture, signal integrity, and environmental noise. As a result, implementations that are theoretically secure can still be vulnerable when deployed on real devices. Evaluating and mitigating side-channel leakage \cite{gan2024classic} is therefore essential, particularly for embedded and resource-constrained platforms, where limited countermeasures and noisy measurement conditions can significantly affect the security of cryptographic operations.

Information Set Decoding (ISD) \cite{peters2010information} belongs to the aforementioned structural attacks and can be augmented by SCA \cite{colombier2022profiled} significantly reducing the practical complexity of algebraic cryptanalysis. Although classical ISD treats all candidate information sets equally, side-channel observations can introduce measurable biases that guide the selection or pruning of candidate information subsets to perform ISD attack. By correlating leakage patterns with intermediate decoding operations, an attacker can prioritize high-probability information subsets, and thereby shrink the effective search space without modifying the underlying code structure.

%
%
\section {Experiment Setup}\label{sec:Experiment}
In the context of SPA against a McEliece cryptosystem \cite{molter2011simple}, this paper attempts to correlate recovered power traces with intentionally stored secret bytes while specific hardware executes cryptographic operations in software implementations of McEliece and BIKE, two code-based cryptosystems of the Open Quantum Safe (OQS) library.  In general, SPA requires specific experimental equipment to attempt more advanced side-channel techniques, as it relies on indirect observation of power consumption patterns followed by extensive statistical processing. Essential tools include a target device that performs cryptographic operations, a stable power supply, and a means of measuring instantaneous power consumption, such as a shunt resistor or current probe coupled with a high-bandwidth digital oscilloscope. A trigger mechanism is often used to align measurements with specific execution phases, while basic data acquisition software enables visualization and manual inspection of traces. Unlike differential techniques, SPA generally does not require large trace sets or complex analysis frameworks, making it accessible yet effective against implementations with clearly distinguishable power signatures.

\begin{figure}[H]
    \centering
    \includegraphics[width=1\linewidth]{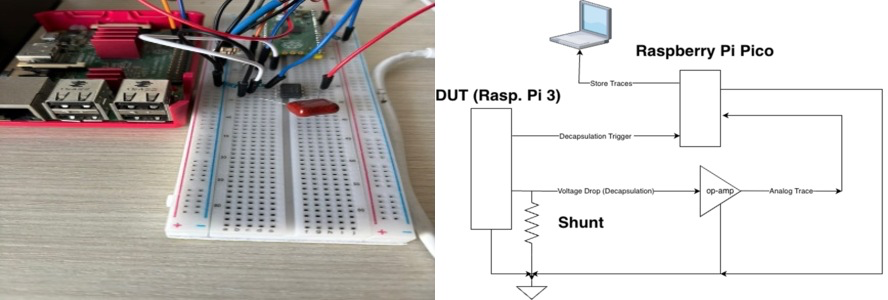}
    \caption{Experiment Equipment Setup and Connections.}
    \label{fig:setup}
\end{figure}

In line with this rationale, the low cost equipment that was recruited to participate in such an experiment, aligned with the setup in \cite{edgeimpulseCollectData}, acting analogously with a realistic SPA one, is the following: 1)A raspberry pi 3 that acts as a device under test (DUT). The DUT executes the cryptographic operations while exposing physical characteristics, such as power consumption or electromagnetic emissions, that may unintentionally leak information. 2)A shunt resistor, which is a low‐resistance component that observes the voltage drop during DUT's cryptographic operations. 3) An operational amplifier (op‐amp) is used to amplify the small voltage drosp across the shunt resistor. 4) A Raspberry pi Pico, which is a low‐cost microcontroller platform used to translate and store op‐amp voltage variations by shunt, monitoring cryptographic operations. Having this set up, schematically represented in Fig. \ref{fig:setup} , it is possible to create an artificial, yet realistic, SCA experiment. This kind of setup has a significant trade‐off, such as low  sampling resolution which  may also include noise.

For this experiment, the Python wrapper of the OQS open source library was used, where the code based Key Encapsulation Mechanisms (KEMs), McEliece and BIKE, both belong to the National Institute of Standards and Technology (NIST) security level‐3. These algorithms run inside the DUT and each captured trace corresponds to a different decapsulation phase where the shared secret (ephemeral) key, is generated.

%

%
%
\section{Results Evaluation}\label{sec:Results}

After executing the experiment, 250 traces were collected for each cryptosystem. Each trace contains 1500 voltage samples produced during the decapsulation phase, also used to construct the following voltage plots.  To evaluate these observations of the experiment, the statistical method of null hypothesis \cite{hussey2023systematic} is used, in addition to the correlation between the voltage drop measurements and some bits of the shared secret key, which were produced and intentionally stored, during decapsulation. Hence, the correlation coefficient $\rho = \frac{\sum (x-\overline{x})(y-\overline{y})}{\sum \sqrt{(x-\overline{x})^2}\sqrt{(y-\overline{y})^2}}$ is computed, where $x$ is the voltage power detected by the shunt and stored in the pi micro-controller and $y$ is the Ηamming weight of the last byte of the shared secret key. The values of $\rho$ exist in $[-1,1]$, where $\rho=0$ indicates that there is no correlation, on the other hand, values such as $-1$ or $1$ indicate absolute negative or positive correlation between quantities $x$ and $y$, respectively.

\begin{figure}[H]
    \centering
    \includegraphics[width=1 \linewidth]{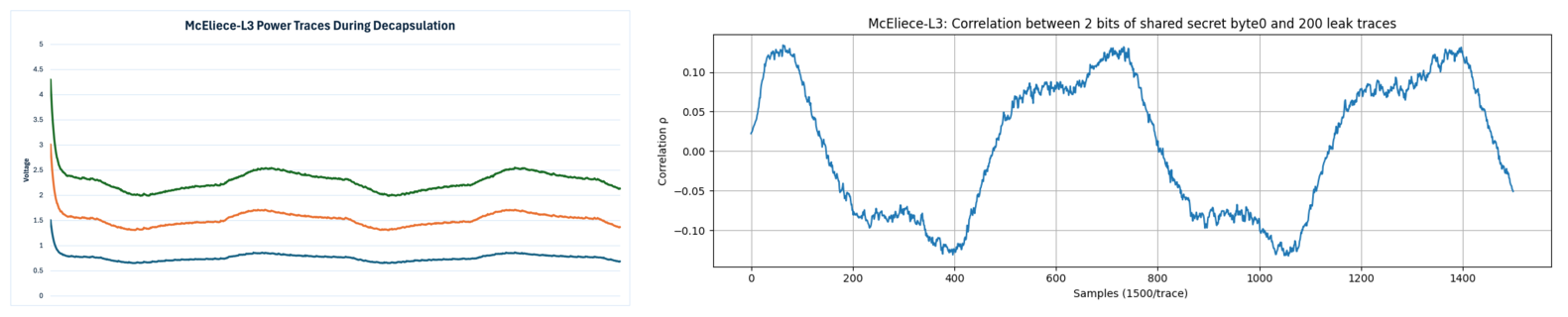}
    \caption{McEliece Power Traces and Correlation Coefficient Graphs. }
    \label{fig:McEliece}
\end{figure}

\begin{figure}[H]
    \centering
    \includegraphics[width=1 \linewidth]{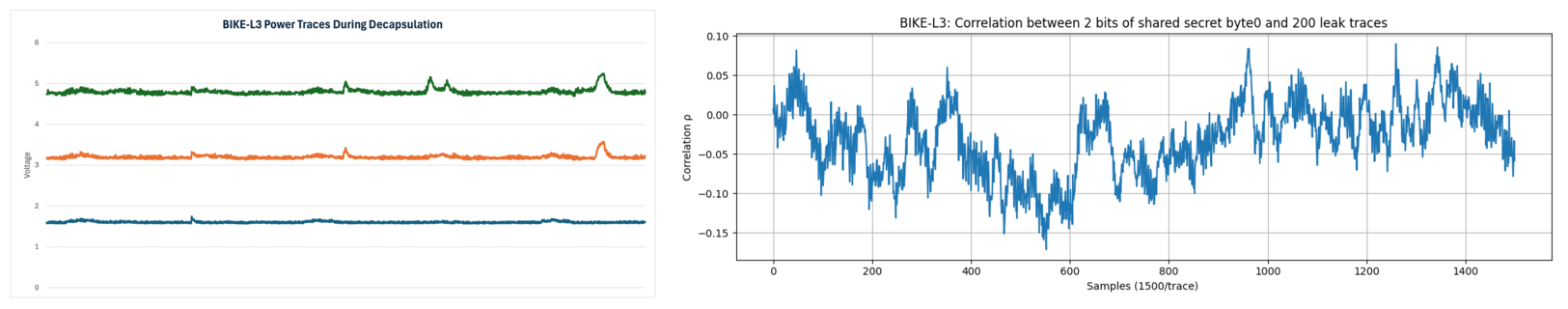}
    \caption{BIKE Power Traces and Correlation Coefficient Graphs. }
    \label{fig:BIKE}
\end{figure}

Declaring the null hypothesis statement ”\textit{Software implementations of code-based post-quantum cryptography do not leak information}”, the correlation $\rho$ for McEliece and BIKE is calculated for analysis. Based on the literature that values of $\rho>0.1$ indicate leakage \cite{bottinelli2017computational}, the initial hypothesis is rejected, leading to the conclusion that these algorithms leak significant information. For example, Fig. \ref{fig:McEliece} represents how, during the decapsulation of the shared secret key from the McEliece KEM, the voltage spikes in the left plot correspond to high variations of $\rho$. The results also demonstrate that in the final phase of McEliece decapsulation there is measurable and temporally aligned power leakage that can be exploited through correlation based SPA. In addition, for the BIKE KEM, the results in Fig. \ref{fig:BIKE} show a correlation exceedance beyond the threshold of $\rho=0.1$, implying that there is more correlation between the traces and the ephemeral key. All these observations allow for further analysis, such as the prediction of secret bits using Machine Learning (ML) models.  

To highlight the above conclusions, some well known classifiers were recruited to evaluate the accuracy in predicting valuable secret information. To train the models, the samples from each trace were used as the features X, while the secret bits from the shared session key were used as the labels Y. The models were configured to use the $80\%$ of the traces dataset for training and the rest for testing their performance. After running the models, the produced data are represented in Tab. \ref{tab:ML}. In particular, we can see that there is a notable probability of predicting the bit $1$, which is higher than a raw guess. In addition, bit $1$ is more likely to be predicted than bit $0$ due to higher voltage spikes that are equivalent to the logical bit $1$. In general, the McEliece cryptosystem appears to leak scalable information, where the F1 score is constantly greater than $0.65$. Note that considering the probability of simply guessing the value of a bit, all the classifiers  appear to  perform $25\%$ better than the simple guess. As far as predicting more than one bit, the models behave in the same manner. For example, in terms of attacking the McEliece, at the same Tab. \ref{tab:ML} we can observe that the prediction of two bits, such as the $10$, is at least $40\%$ greater than the simple guess, namely $25\%$.
\begin{table}[H]
\centering

\label{tab:ML}
\begin{tabular}{lccc}
\hline
\textbf{Classifier} & \textbf{Class bit 0 F1} & \textbf{Class bit 1 F1} & \textbf{Accuracy} \\
\hline
kNN(vs McEliece)        & 0.57 & 0.67 & 0.63\\
Random Forest(vs BIKE)        & 0.53 & \textbf{0.65} & 0.60\\
Linear SVM(vs McEliece/BIKE) & 0.47/0.60 & \textbf{0.68}/0.60 & 0.60/0.60 \\
\hline
\textbf{} & \textbf{Recall} & \textbf{ F1} & \textbf{Accuracy} \\
\hline
Linear SVM(vs McEliece bits 10) & 0.43 & 0.35 & 0.33 \\
\hline
\end{tabular}
\caption{ Bit Prediction Using Only 250 Traces During Decapsulation.}
\end{table}

%
%
\section{Conclusions and Future Work}
The null hypothesis, along with the SPA, was used to evaluate electromagnetic emissions of code-based post-quantum cryptosystems utilizing trivial equipment with low resolution capabilities. The results indicate that there is a correlation between the captured power traces and secret bits. Variations in the observed waveforms are consistent with the predicted leakage model, suggesting the presence of secret‐dependent information. To mitigate these phenomena, several countermeasures have been studied, focusing on masking the cryptographic operation during the decapsulation. Hence, performing an attack based on SPA, using the paper's low cost setup, would be a challenging effort to evaluate the masking countermeasures. Furthermore, the Hamming Quasi-Cyclic (HQC) KEM, selected by the NIST as another quantum resistant code-based standard mechanism, was not included in the current experiment because it is not part of the OQS, and consequently, suggesting a promising study.

\printbibliography

\end{document}